\newread\epsffilein    
\newif\ifepsffileok    
\newif\ifepsfbbfound   
\newif\ifepsfverbose   
\newdimen\epsfxsize    
\newdimen\epsfysize    
\newdimen\epsftsize    
\newdimen\epsfrsize    
\newdimen\epsftmp      
\newdimen\pspoints     
\def\epsfbox#1{\global\def\epsfllx{72}\global\def\epsflly{72}%
   \global\def\epsfurx{540}\global\def\epsfury{720}%
   \def\lbracket{[}\def\testit{#1}\ifx\testit\lbracket
   \let\next=\epsfgetlitbb\else\let\next=\epsfnormal\fi\next{#1}}%
\def\epsfgetlitbb#1#2 #3 #4 #5]#6{\epsfgrab #2 #3 #4 #5 .\\%
   \epsfsetgraph{#6}}%
\def\epsfnormal#1{\epsfgetbb{#1}\epsfsetgraph{#1}}%
\def\epsfgetbb#1{%
\openin\epsffilein=#1
\ifeof\epsffilein\errmessage{I couldn't open #1, will ignore it}\else
   {\epsffileoktrue \chardef\other=12
    \def\do##1{\catcode`##1=\other}\dospecials \catcode`\ =10
    \loop
       \read\epsffilein to \epsffileline
       \ifeof\epsffilein\epsffileokfalse\else
          \expandafter\epsfaux\epsffileline:. \\%
       \fi
   \ifepsffileok\repeat
   \ifepsfbbfound\else
    \ifepsfverbose\message{No bounding box comment in #1; using defaults}\fi\fi
   }\closein\epsffilein\fi}%
\def\epsfsetgraph#1{%
   \epsfrsize=\epsfury\pspoints
   \advance\epsfrsize by-\epsflly\pspoints
   \epsftsize=\epsfurx\pspoints
   \advance\epsftsize by-\epsfllx\pspoints
   \epsfxsize\epsfsize\epsftsize\epsfrsize
   \ifnum\epsfxsize=0 \ifnum\epsfysize=0
      \epsfxsize=\epsftsize \epsfysize=\epsfrsize
     \else\epsftmp=\epsftsize \divide\epsftmp\epsfrsize
       \epsfxsize=\epsfysize \multiply\epsfxsize\epsftmp
       \multiply\epsftmp\epsfrsize \advance\epsftsize-\epsftmp
       \epsftmp=\epsfysize
       \loop \advance\epsftsize\epsftsize \divide\epsftmp 2
       \ifnum\epsftmp>0
          \ifnum\epsftsize<\epsfrsize\else
             \advance\epsftsize-\epsfrsize \advance\epsfxsize\epsftmp \fi
       \repeat
     \fi
   \else\epsftmp=\epsfrsize \divide\epsftmp\epsftsize
     \epsfysize=\epsfxsize \multiply\epsfysize\epsftmp   
     \multiply\epsftmp\epsftsize \advance\epsfrsize-\epsftmp
     \epsftmp=\epsfxsize
     \loop \advance\epsfrsize\epsfrsize \divide\epsftmp 2
     \ifnum\epsftmp>0
        \ifnum\epsfrsize<\epsftsize\else
           \advance\epsfrsize-\epsftsize \advance\epsfysize\epsftmp \fi
     \repeat     
   \fi
   \ifepsfverbose\message{#1: width=\the\epsfxsize, height=\the\epsfysize}\fi
   \epsftmp=10\epsfxsize \divide\epsftmp\pspoints
   \vbox to\epsfysize{\vfil\hbox to\epsfxsize{%
      \includegraphics{#1}%
      \hfil}}%
\epsfxsize=0pt\epsfysize=0pt}%

{\catcode`\%=12 \global\let\epsfpercent=
\long\def\epsfaux#1#2:#3\\{\ifx#1\epsfpercent
   \def\testit{#2}\ifx\testit\epsfbblit
      \epsfgrab #3 . . . \\%
      \epsffileokfalse
      \global\epsfbbfoundtrue
   \fi\else\ifx#1\par\else\epsffileokfalse\fi\fi}%
\def\epsfgrab #1 #2 #3 #4 #5\\{%
   \global\def\epsfllx{#1}\ifx\epsfllx\empty
      \epsfgrab #2 #3 #4 #5 .\\\else
   \global\def\epsflly{#2}%
   \global\def\epsfurx{#3}\global\def\epsfury{#4}\fi}%
\def\epsfsize#1#2{\epsfxsize}
\let\epsffile=\epsfbox

\documentstyle[11pt,paspconf]{article}

\markboth{Biemes, et al.}{Test paper}

\begin{document}

\title{Collisionless Relaxation of Stellar Systems}

\author{Henry E. Kandrup}
\affil{Department of Astronomy and Department of Physics and 
Institute for Fundamental Theory, University of Florida, Gainesvile, FL 32611}

\begin{abstract}
The objective of the work summarised here has been to exploit and extend ideas
from plasma physics and accelerator dynamics to formulate a unified description
of collisionless relaxation that views violent relaxation, Landau damping, and
phase mixing as (manifestations of) a single phenomenon. This approach embraces
the fact that the collisionless Boltzmann equation ({\it CBE}), the basic 
object of the theory, is an infinite-dimensional Hamiltonian system, with
the distribution function $f$ playing the role of the fundamental dynamical
variable, and that, interpreted appropriately, an evolution described by the 
{\it CBE} is no different fundamentally from an evolution described by any 
other Hamiltonian system. Equilibrium solutions $f_{0}$ correspond to extremal 
points of the Hamiltonian subject to the constraints associated with 
Liouville's Theorem. Stable equilibria correspond to energy minima. The 
evolution of a system out of equilibrium involves (in general nonlinear) phase 
space oscillations which may -- or may not -- interfere destructively so as to 
damp away. 
\end{abstract}
\keywords{galaxies,kinematics and dynamics,evolution}
\vskip .3in
\section{Introduction}
A satisfactory theory of collisionless relaxation must address two general 
issues, namely (1) the form of the equilibrium towards which any given set of 
initial conditions should evolve and (2) the overall efficiency of this 
approach towards equilibrium or, more generally, the physics that drives the 
evolution. For the case of real galaxies, these questions will be impacted to 
at least some extent by dissipation associated, e.g., with gas dynamics and/or 
discreteness effects, i.e., collisionality. Collisionless relaxation is an 
idealisation in which these effects are completely ignored.

Work in this area dates back at least to the 1960's, when a number
of different workers, including H\'enon (1961), King (1962), and Lynden-Bell
(1967) argued (1) that there should be a coarse-grained approach {\it towards} 
(albeit not necessarily {\it to}) a unique equilbriium, namely that described 
by so-called Lynden-Bell statistics, and (2) that this approach should proceed 
on a dynamical time scale, $t_{D}$, since it is a collective process, unlike 
the theory of collisional relaxation which had been developed by Chandrasekhar 
some twenty years earlier. Later workers subsequently shifted the focus in
a number of different ways, notably by asking specifically which initial
conditions give rise to which final equilibria (e.g., Ziegler \& Wiechen 
1989). However, it is probably fair to say that almost all the work on 
collisionless relaxation over the past thirty years has been formulated in the 
context of ideas that evolved in the 1960's.

The objective of the work described here has been to apply to collisionless
self-gravitating systems various ideas and techniques from plasma physics and 
accelerator dynamics which have proven successful in explaining real 
experiments, the obvious point being that, unlike galactic dynamics, these 
disciplines allow for the possibility of controlled expriments whereby 
theoretical predictions can be confirmed and/or falsified. Section 2 discusses 
some issues related to the interpretation of the {\it CBE}, the starting point
for any theory of collisionless relaxation. Section 3 then identifies the 
correct mathematical sense in which, as is generally asserted, the {\it CBE} 
is a Hamiltonian system. Section 4 exploits this Hamiltonian formulation to 
present a coherent description of what happens as a generic initial condition 
evolves into the future. Section 5 concludes with a discussion of the phase 
mixing exhibited by {\it CBE} characteristics, i.e., orbits in the 
self-consistent potential, focusing in particular on the possible role of 
chaos. 

\section{The Collisionless Boltzmann Equation}

The basic assumption underlying collisionless relaxation is that the
system in question can be described by a one-particle distribution function, 
$f({\bf x},{\bf v},t)$, defined as a phase space mass density, the evolution of
which is governed by the collisionless Boltzmann equation ({\it CBE}), which
implies free streaming in the self-consistent potential associated with $f$
(H\'enon 1982).

It is often asserted that, in the $N\to\infty$ limit, where collisionality 
becomes irrelevant, the {\it CBE} and the $N$-body problem coincide. However, 
this correspondence is not completely trivial. For the special case of a 
singular initial condition, 
\begin{equation}
f({\bf x},{\bf v},0)=\sum_{i=1}^{N}\,m_{i}\,
{\delta}_{D}[{\bf x}-{\bf x}_{i}(0)] {\delta}_{D}[{\bf v}-{\bf v}_{i}(0)], 
\end{equation}
in which point masses are located in phase space with arbitrary precision,
a solution to the {\it CBE} is equivalent mathematically to the full
$N$-body problem. However, such singular solutions are not what interest the 
theorist exploiting the {\it CBE}. Rather, he or she is interested typically 
in the evolution of a smooth initial $f(0)$ or the construction of a 
smooth equilibrium $f_{0}$ which may be given as a function of global isolating
integrals (although this is not necessary (Kandrup 1998c)). A crucial piece of
the physics thus entails passing from singular to smooth distribution 
functions, which is decidedly nontrivial. Perhaps the best way to give meaning
to such a smooth $f$ is to adopt the tact taken by plasma physicists (e.g.,
Klimontovich 1980) and cosmologists (Peebles 1980), assuming that any given
realisation of the $N$-body problem, performed either by nature or a computer,
entails sampling a smooth $f(0)$ to generate initial conditions for the 
$N$-body problem and then evolving these into the future.

However, given this interpretation there is no reason {\it a priori} to 
expect smooth {\it CBE} characteristics, i.e., orbits in the smooth 
potential associated with $f$, to have anything to do with $N$-body orbits.
In particular, there is no contradiction between $N$-body orbits which are 
chaotic and {\it CBE} characteristics which are completely integrable. 
This is well known from ordinary kinetic theory. One anticipates that orbits
in a gas of molecules interacting via short range forces will be chaotic
(as has been proved for the case of hard sphere interactions), but the mean
field characteristics for a homogeneous system in equilibrium are all 
integrable constant velocity trajectories.

Nevertheless, much of the galactic dynamicist's intuition is based on orbits
in smooth potentials, so that one might hope that there is some average sense 
in which, at least for finite time intervals, {\it CBE} characteristics track
$N$-body orbits. No hard results about this issue have yet been proved.
However, there is at least one conjecture (Kandrup 1998a) which has not yet 
been disproved: Suppose that one performs an ensemble of different $N$-body
simulations, all generated by sampling the same smooth $f(0)$ but with 
one orbit always starting at the same $({\bf r}(0),{\bf v}(0))$. One might
then suppose that, at least for large $N$, the {\it rms} deviation of the
$N$-body orbits from the {\it CBE} characteristic associated with the same
initial $({\bf r}(0),{\bf v}(0))$ will satisfy
\begin{equation}
{\delta}r_{rms}(t){\;}{\sim}{\;}F(N){\exp}(t/{\tau}),
\end{equation}
where ${\tau}{\;}{\sim}{\;}t_{D}$ is the characteristic Lyapunov time on which
solutions to the $N$-body problem diverge, and $F(N)$ is a decreasing function 
of $N$ (see, e.g., Kandrup \& Smith 1991, Goodman, Heggie, \& Hut 1994).

\section{The Hamiltonian Character of the Evolution}

Galactic astronomers are cognizant of the fact that the {\it CBE} is a
Hamiltonian system. However, most probably do not know the exact sense in
which this is true. There is the implicit idea that, because {\it CBE} 
characteristics correspond to orbits in a (in general) time-dependent
Hamiltonian system, the {\it CBE} must itself be Hamiltonian. However, this is
not really the point. Viewing the physics as corresponding to orbits in a
time-dependent potential is a cheat since this does not incorporate 
self-consistency in a fundamental way! Rather, a proper Hamiltonian formulation
entails a very different viewpoint in which the distribution function $f$ 
itself is the fundamental dynamical variable. 

Such a Hamiltonian formulation was first presented by Morrison (1980) for the 
Vlasov equation, the plasma analogue of the {\it CBE}. Here $f$ is taken as 
the basic dynamical variable which evolves in the infinite-dimensional phase 
space of distribution functions. For the case of gravitational interactions, 
the Hamiltonian 
$$\hskip -2in {\cal H}[f]=
\,\int\,d^{3}xd^{3}v\,{\small{1\over 2}}\,v^{2}\,f({\bf x},{\bf v},t)$$
\begin{equation}
-{G\over 2}\;\int\,d^{3}x\,d^{3}v\;\int\,d^{2}x'\,d^{3}v'\;
{f({\bf x},{\bf v},t)f({\bf x}',{\bf v}',t)\over |{\bf x}-{\bf x}'|} 
\end{equation}
is nothing other than the mean field energy, as identified, e.g., by 
Lynden-Bell \& Sanitt 1969). The first term represents the mean kinetic energy 
and the second represents the gravitational potential energy associated with
the mass density ${\rho}=\int\,d^{3}v\,f$. The {\it CBE} is most 
easily written in a Hamiltonian form by generalising the notion of an ordinary
Poisson bracket. Specifically, given two functionals of the distribution 
function $f$, say ${\cal A}[f]$ and ${\cal B}[f]$, one can define the action of
a Lie bracket $[\, . \, , \, . \,]$ as 
\begin{equation}
 [{\cal A},{\cal B}]=\int d^{3}xd^{3}v\,f\,{\Bigl\{}
{{\delta}{\cal A}\over {\delta}f},{{\delta}{\cal B}\over {\delta}f}{\Bigr\}} 
\end{equation}
where $\{ \; .\; ,\; .\; \}$ is the the ordinary Poisson bracket and 
${\delta}/{\delta}f$ is a functional derivative. What is important about this
Lie bracket is that, like the Poisson bracket, it is a linear
antisymmetric operation that satisfies the Jacobi identity, which implies
that it can be used to generate a Hamiltonian evolution (see, e,g,, Arnold
1989). The key point then is that, in terms of the Hamiltonian (3), the 
Lie bracket (4) yields the {\it CBE} as
\begin{equation}
{{\partial}f\over {\partial}t}+[{\cal H},f] = 0 .
\end{equation}
By exploiting the fact that 
\begin{equation}
{\delta}{\cal H}/{\delta}f = E={1\over 2}v^{2}+{\Phi}({\bf x},t), 
\end{equation}
where $E$ is the energy of a test particle moving in the self-consistent
potential ${\Phi}[f]$, eq. (5) can be rewritten in the standard Poisson bracket
form (see, e.g., Binney and Tremaine 1987)
\begin{equation}
{{\partial}f\over {\partial}t}- \{ E,f \} =0. 
\end{equation}

One important feature associated with the {\it CBE} is Liouville's Theorem,
which implies the existence of an infinite number of conserved quantities,
the so-called Casimirs. Specifically, any function ${\chi}(f)$ defines a phase 
space functional
\begin{equation}
C[f]=\int d^{3}xd^{3}v\;{\chi}(f) , 
\end{equation}
the numerical value of which is constant in time, i.e., 
$dC/dt {\;}{\equiv}{\;} 0$. The fact that $f$ satisfies these constraints,
which are associated with various internal symmetries (Morrison and Eliezur
1996), implies that its evolution is restricted to a reduced, but still 
infinite-dimensional, hypersurface in the phase space of distribution 
functions. 

All of this is correct mathematically, but still one might ask: what are the
$p$'s and $q$'s? What are the canonically conjugate variables corresponding to
the coordinates and momenta of ordinary particle mechanics? The short answer
to this is that, as for other Hamiltonian theories of continuous media, such
as the equations for an incompressible two-dimensional fluid, there is no
easy decomposition of the dynamical variables into canonical pairs (see, e.g.,
Morrison 1998). Once one has restricted attention to a reduced phase space in 
which the values of all the Casimirs are fixed, Darboux's Theorem (see, e.g.,
Arnold 1989) implies that, at least locally, such conjugate variables exist. 
However, identifying them is hard and, moreover, there is no guarantee that 
they can smoothly cover the entire phase space. This means that, when 
visualising an evolution governed by the {\it CBE}, one {\it can} view the 
physics as being no different fundamentally from more familiar Hamiltonian 
systems, but that one has to visualise the evolution in phase space, rather 
than configuration space. Given a ``feel'' for Hamiltonian mechanics as viewed 
in phase space, and an appreciation of the subtleties that arise when one 
allows for infinite degrees of freedom (in particular, the notion of phase 
mixing discussed in the following section), visualising an evolution governed 
by the {\it CBE} is really not all that complicated.

To make all this mathematics somewhat less obscure, it is useful to consider a 
much simpler mechanics problem that incorporates many of the same features, 
namely the equations of motion for a freely rotating three-dimensional rigid 
body. Specifically, the Euler equations for solid body rotations constitute 
(cf. Kandrup 1990) a non-canonical Hamiltonian system with a three-dimensional
phase space and an evolution generated by a Lie bracket that is {\it not} the
Poisson bracket of ordinary particle mechanics.

Working in Cartesian coordinates, the three dynamical variables can be taken 
as the three components of angular momentum, $L_{i}$ ($i=x,y,z$), which 
coordinatize a three-dimensional phase space. The Hamiltonian is nothing other
than the kinetic energy which, in terms of the principal moments of inertia, 
takes the form
\begin{equation}
H=\sum_{i=1}^3 {L_{i}^{2}\over 2I_{i}}.
\end{equation}
The Lie bracket corresponds to the generator of the three-dimensional 
rotation group, for which
\begin{equation}
[A,B]=\sum_{i}\sum_{j}\sum_{k}
{\epsilon}_{ijk}L_k{{\partial}A\over {\partial}L_{i}}
{{\partial}B\over {\partial}L_{j}}{\;}{\equiv}{\;}\sum_{i}\sum_{j}\sum_{k}
J_{ij}{{\partial}A\over {\partial}L_{i}}{{\partial}B\over {\partial}L_{j}}.
\end{equation}
It is easily verified that this Hamiltonian and this bracket combine to yield
equations of motion of the form
\begin{equation}
\dot L_i=[L_{i},H]
=\sum_{j}\sum_{k}{\epsilon}_{ijk}L_{k}L_{j} I_{j}^{-1}, \qquad (i=x,y,z) 
\end{equation}
which are nothing other than the ordinary Euler equations.

An evolution governed by these equations is constrained by the existence
of a conserved quantity, namely the total squared angular momentum 
\begin{equation}
C[L_{i}]={\,}\sum_{i=1}^3L_{i}^{2}. 
\end{equation}
It follows that only two of the three componnets of angular momentum can be 
specified independently, and that the effective phase space is really 
two-dimensional. In principle, one can extract a canonically 
conjugate pair of variables, but doing so is not all that easy (or 
illuminating). 

\section{The Approach towards Equilibrium}

Stated suscinctly, an evolution governed by the {\it CBE} corresponds to
oscillations in an infinite-dimensional phase space, oscillations which, in
many cases, may exhibit destructive interference and, consequently, damp away:
An infinitesimal perturbation away from a stable equilibrium $f_{0}$ will, 
when evolved into the future, exhibit linear phase space oscillations which 
may or may not exhibit Landau damping/phase mixing. Nonlinear deviations will 
exhibit nonlinear oscillations about one or more equilibria. 
Because the {\it CBE} is Hamiltonian, an initial $f(0)$ cannot exhibit a
pointwise approach towards an equilibrium distribution $f_{0}$. However, one
{\it can} get a phase mixing of (linear or nonlinear) modes which implies 
that, as probed by the behaviour of such observables as the mass density 
${\rho}({\bf x})$ or the velocity dispersion tensor ${\sigma}_{ij}({\bf x})$, 
there is an approach towards a time-independent $f_{0}$ (Kandrup 1998b).

The obvious question, of course, is: towards which equilibrium $f_{0}$ will
some initial $f(0)$ evolve? The key to addressing this issue is the
recognition that the {\it CBE} implies that all equilibrium solutions $f_{0}$
are energy extremals. More precisely, one can prove that a distribution 
function $f_{0}$ corresponds to a time-independent solution, for which 
${\partial}f_{0}/{\partial}t{\;}{\equiv}{\;}0$, if and only if the first 
variation of the Hamiltonian (3) vanishes identically for all perturbations
${\delta}f$ that satisfy the constraints associated with Liouville's Theorem,
i.e., ${\delta}^{(1)}{\cal H}{\;}{\equiv}{\;}0.$ This result, which was 
first discovered in the context of galactic dynamics by Bartholomew (1971)
and subsequently discovered by the plasma physicists nearly two decades 
later, follows as a direct corollary of the observation that the ``dynamically
accessible'' perturbations which satisfy all the constraints associated with
Liouville's Theorem are all related to the original $f_{0}$ by a canonical
transformation in terms of some generating function $h$, i.e., 
\begin{equation}
f_{0} \to e^{\{h,\, . \,\}} f_{0} = f_{0} + \{h,f_{0}\} + 
{1\over 2!}{\bigl\{} h, \{ h,f_{0} \} {\bigr\}} + ....
\end{equation}

It is easy to see that, if the energy extremal is a local minimum, so that
the second variation ${\delta}^{(2)}{\cal H}{\;}{\ge}{\;}0$ for all 
linear perturbations, the equilibrium is linearly stable. When subjected to a
linear perturbation, the distribution function will execute phase space 
motions analogous to a particle which, initially at 
rest at the minimum of some potential, is given an infinitesimal phase space 
perturbation corresponding to a nonnegative kinetic energy and
a nonnegative potential energy relative to the extremal point.

Far less trivial, but also true, is the fact that, for fixed values of all 
the Casimirs $C[f]$, there is always a global energy minimum. This was 
first proven directly by Wiechen, Ziegler, \& Schindler (1988). Alternatively, 
the existence of a global minimum follows as a corollary to the proof of 
global existence first established by Pfaffelmoser (1992). Global existence, 
i.e., the fact that sufficiently smooth initial data never give rise to such
singular behaviour as caustics or shocks, is itself important, and not 
completely obvious physically. One knows, for example, that global existence 
does {\it not} hold for the equations that describe a perfect fluid, even 
though these equations can be derived from the {\it CBE} by implementing a 
truncated moment expansion.

If, for fixed values of the conserved quantities $C[f]$, there exists only
one energy extremal, $f_{0}$, that extremal must correspond to a global energy 
minimum and $f_{0}$ must be globally stable. This implies that {\it any} 
$f(0)$ generated from $f_{0}$ by a canonical transformation which leaves 
invariant all the constraints associated with Liouville's Theorem will simply
evolve so as to execute (in general nonlinear) oscillations about the minimum
energy state. Explicit examples of such globally stable equilibria, which 
include some of the spherical polytropes, have been constructed by Aly (see,
e.g., Aly 1994).

In a more general setting, where the reduced phase space hypersurface admits
more than  one energy extremal, the physics is more complicated in practice 
although what is going on is still straightforward in principle. In such 
settings, the distribution function $f$ will in oscillate about one or more 
energy minima just like a particle in a potential can oscillate about
one or more energy minima, each of which corresponds to an extremal point 
where the kinetic energy vanishes and the potential energy is a local minimum.

But how should one visualise the evolution of a small perturbation away from a 
collisionless equilibrium? Here the important message is that one ought not 
to distinguish between Landau damping and phase mixing: Viewed appropriately,
Landau damping {\it is} the phase mixing of a wave packet of normal modes. This
was first shown for the case of a homogeneous electrostatic plasma by Case 
(1959), who first computed a complete set of normal modes for the evolution 
equation satisfied by a linearised perturbation and then demonstrated that 
wave packets constructed from these modes damp at the rate that Landau (1946) 
had originally derived.

It is not, however, true that every linearised perturbation of every 
collisionless equilibrium will damp! Linear Landau damping is guaranteed if
all the normal modes are continuous (i.e., if the linearised evolution equation
has no point spectrum), but discrete modes correspond to physical perturbations
that need not damp away (Habib, Kandrup, \& Yip 1986). As discussed more 
carefully in Kandrup (1988b), the physics here is analogous to what arises in
ordinary quantum mechanics. If one considers an observable like angular 
momentum with a discrete spectrum, it is possible to construct well behaved
eigenstates which, when evolved into the future, maintain their coherence for
all times. If, however, one considers observables like position or linear
momentum, where the spectrum is continuous, the situation is completely 
different. In this case, a nonsingular initial state must be constructed from
a continuous set of singular eigendistribution, so that the best that one
can do is construct a localised wave packet. However, when evolved into the 
future such a wavepacket will necessarily spread because different 
eigendistributions have different phase velocities.

Especially given the common intuition that perturbations of a ``realistic'' 
plasma will always Landau damp, it is important to stress that there {\it are}
known examples, both in plasma physics and galactic dynamics, of geometries
where one can have solutions characterised by undamped oscillations.
Perhaps the best known example is provided by the so-called Van Kampen (1955)
modes, which arise oftentimes in a plasma when the equilibrium distribution 
vanishes identically for particle speeds larger than some critical
value. First predicted analytically, these modes are well known to 
experimentalists. 

Louis \& Gerhart (1988) and Sridhar (1989) have demonstrated explicitly that
one can also construct models of spherical galaxies that correspond seemingly
to time-independent equilibria perturbed by finite amplitude undamped 
oscillations. Whether such configurations could arise as a consequence of a
realistic, or quasi-realistic, evolution is not clear. However, numerical 
simulations by Mineau et al (1990) have shown that, at least for the toy model
of one-dimensional gravity with infinite plane sheets, it is possible to 
create such a pulsating configuration by colliding two reasonable 
time-independent equilibria. These gravitational examples are more complicated 
than the simple examples that give rise to Van Kampen modes, but the physics 
is very similar. 

The conventional picture of Landau damping (see, e.g., Stix 1962) interprets 
the damping as resulting from a resonance between unperturbed particles
moving with physical velocity {\bf v} and a wave, i.e., the perturbation,
propagating with a phase velocity ${\bf c}$. The obvious point, then, is that
if there are no particles with physical velocity ${\bf c}$, a wave travelling 
with velocity ${\bf c}$ has nothing with which to resonate so that it will
not damp. Landau's original derivation of damping via an evaluation of poles 
in the complex plane is only valid for equilibria that are analytic functions 
of velocity, which precludes the possibility of velocities for which the 
unperturbed $f_{0}$ vanishes identically. 

Modulo nontrivial boundary conditions such as phase space holes, linearised 
perturbations of a stable equilibrium would be expected to exhibit linear 
Landau damping; and, by analogy, one might expect that larger amplitude
perturbations will exhibit nonlinear Landau damping. However, if one visualises
a generic initial $f(0)$ as a large perturbation away from {\it some} e
quilibrium, the form of which one need not know, it would appear reasonable to 
view its possible approach towards equilibrium as a manifestation of nonlinear 
Landau damping. It thus seems natural to argue that the phenomenon that the 
astronomer is wont to interpret as violent relaxation is really the 
gravitational analogue of what the plasma physicist terms nonlinear Landau 
damping.

One potential advantage of such an interpretation is the fact that the words
Landau damping suggest an evolution that exhibits significantly more coherence
than is usually assumed in violent relaxation. Lynden-Bell's (1967)
original analysis presupposed that the coarse-grained final equilibrium is
``random,'' subject only to some set of holonomic constraints (including 
conservation of number and energy) and a coarse-grained version of Liouville's
Theorem, which ensures that the phase space evolution is incompressible. 
However, numerical simulations by numerous workers (e.g., Van Albada 1982,
Quinn \& Zurek 1988, Kandrup et al 1994), even those with
almost no softening, demonstrate unambiguously that individual particles
(and thus, presumably, phase elements in the {\it CBE}) exhibit a significant
remembrance of things past in terms of their binding energies.

\section{Regular and Chaotic Phase Mixing}

Section 4 focused on the phase mixing exhibited by $f(t)$ as it evolves in the 
phase space of distribution functions. Also interesting, however, and more 
consonant with the usual way in which galactic astronomers view collisionless 
relaxation, is the question of how initially localised phase space elements 
disperse as a result of an evolution governed by the {\it CBE}. Honest 
computations of this sort have not yet been effected numerically. However, 
some insights into what one might expect can be derived by considering a much 
simpler problem, namely the mixing exhibited by a phase element evolved in a 
fixed potential. 

What happens if an initially localised
phase element (or, equivalently, a collection of points sampling that
phase element) is evolved into the future? How fast, and in what sense, will
one observe an approach towards some time-independent, or nearly 
time-independent equilibrium? Does the form of the evolution depend sensitively
on the details of the potential, or does it depend simply on whether the
phase element corresponds to chaotic or regular motion?

This problem can be, and has been, investigated by (1) selecting a two-
or three-dimensional potential, (2) choosing a small phase space cell of
fixed energy, (3) sampling that cell to get a collection of individual
initial condition, (4) evolving these initial conditions into the future, and
then (5) analysing the statistical properties of the resulting trajectories as 
a function of time $t$ (Kandrup \& Mahon 1994, Mahon et al 1995, Merritt \& 
Valluri 1996, Kandrup 1997). The statistical analysis involved tracking three
different diagnostics: 
\begin{itemize}
\item The convergence of coarse-grained reduced distributions towards
time-independent forms. This involved, e.g., binning the data to extract such
reduced distributions as $f(y,v_{y},t)$ and then determining whether such 
an $f(t)$ will converge towards a time-independent invariant
$f_{eq}(y,v_{y})$. Convergence was defined with respect to the ``norm''
\begin{equation}
||f(Z_{a},Z_{b},t) - f_{eq}(Z_{z},Z_{b})||=
{\biggl(} \sum_{a}\sum_{b}|f(Z_{a},Z_{b},t)-f_{eq}(Z_{a},Z_{b})|^{p} 
{\biggr)}^{1/p},
\end{equation}
for $Z_{a}{\;}{\ne}{\;}Z_{b}=x,y,z,v_{x},v_{y},$ or $v_{z}$ and with $p=1$ or 
$2$.
\item The evolution of moments such as the velocity dispersion 
${\sigma}_{v_{x}}$ which, for an initially localised phase element, start
small but eventually asymptote towards a much larger value.
\item The evolution of other moments such as the mean velocity 
${\langle}v_{x}{\rangle}$ which, in many cases, eventually asymptote towards
zero. 
\end{itemize}

Investigations of motions in a large number of different potentials have
shown that the detailed choice of potential is relatively unimportant, and
that the most important distinction is between what Merritt \& Valluri have
termed {\it regular} and {\it chaotic phase mixing}. For regular phase 
elements, i.e., elements for which the sampled orbits are all regular, the
reduced distributions approach an equilibrium or near-equilibrium as a power 
law in time, i.e.,
\begin{equation}
||f(t)-f_{eq}||{\;}{\sim}{\;}(t/{\tau})^{-p}.
\end{equation}
The growing moments grow as a power law in time until they saturate at a 
constant value, whereas the decreasing moments damp as a power law in time.
Fully chaotic phase elements, i.e., phase elements in two-dimensional 
potentials with one positive Lyapunov exponent and those in three-dimensional
potentials with two positive Lyapunov exponents, evolve very differently.
For such phase elements, the reduced distributions approach a 
(near-)equilibrium exponentially in time, i.e.,
\begin{equation}
||f(t)-f_{eq}||{\;}{\sim}{\;}\exp (-t/{\tau}),
\end{equation}
the growing moments initially grow exponentially, and the decreasing moments
decrease exponnetially. Partially chaotic three-dimensional phase elements,
with only one positive Lyapunov exponent, exhibit exponential evolution in
some directions and power law evolution in others.

In every case, the natural time scale ${\tau}$ is of order a characteristic
dynamical, or crossing, time, but chaotic and regular phase mixing are very
different: Exponential evolution is much more efficient than power law 
evolution. This simple picture is complicated by various technical points, 
such as the fact that phase mixing can proceed at significantly different 
rates in different phase space directions and that chaotic mixing can be 
impeded for surprisingly long times by such obstructions as cantori or an 
Arnold web (see, e.g., Lichtenberg \& Lieberman 1992). However, these are 
relatively minor perturbations on the basic conclusion: {\it In generating a
well-shuffled state, chaotic mixing is far more efficient than is regular 
mixing.}

But what is actually seen if one examines visually the distorted phase element
or the locations of the points that were evolved from a sampling of the
initial $f(0)$? The full four- or six-dimensional phase element 
becomes stretched and distorted in a volume-preserving fashion which, when 
projected onto a two-dimensional plane, gives rise to a complicated pattern
of tendrils and whorls. The Hamiltonian character of the flow implies that
these whorls can never disappear, but the power
associated with these structures cascades down to progressively shorter scales.
Also interesting, and potentially significant, is the fact that even very weak
non-Hamiltonian perturbations, modeled as friction and noise, can be 
surprisingly efficient in ``fuzzing out'' these short range structures, 
thus facilitating a more complete approach towards a near-equilibrium (see,
e.g., Habib, Kandrup, \& Mahon 1997).


The form of the small scale structures that evolve and the relative efficacy 
of regular and chaotic phase mixing are evident from Figures 1 and 2, which
exhibit the $(x,y)$ coordinates associated with $50625$ orbit samplings of 
initially localised chaotic and regular phase elements in the so-called 
dihedral potential (cf. Mahon et al 1995). In each case the initial phase
ement was generated by uniformly sampling a square of side 
${\Delta}y={\Delta}v_{y}=0.2$ in
the $y-v_{y}$ plane, setting $x=0$, and then solving for 
$v_{x}=v_{x}(x,y,v_{y},E)>0$, with $E=1.0$ the energy.

One final point remains to be stressed: {\it A priori}, whether or not the 
{\it CBE} itself is chaotic need have little to do with whether or not 
{\it CBE} characteristics are chaotic! The fact that portions of the flow in
a self-consistent evolution exhibit chaotic mixing does not imply that the
the {\it CBE} is chaotic. Saying that the {\it CBE} is chaotic should 
be a statement about the evolution of the distribution function in phase space,
not about orbits in the self-consistent potential ${\Phi}$ associated with $f$.

\acknowledgments
It is a pleasure to acknowledge useful discussions, and arguments, with
Paul Channell, Barbara Eckstein, Salman Habib, Ted Kirkpatrick, Donald 
Lynden-Bell, Elaine Mahon, Dave Merritt, Phil Morrison, Tom O'Neil, Daniel 
Pfenniger, Ilya Pogorelov, Haywood Smith, and Christos Siopis.

\vfill\eject
\pagestyle{empty}
\begin{figure}[t]
\centering
\centerline{
        \epsfxsize=9cm
        \epsffile{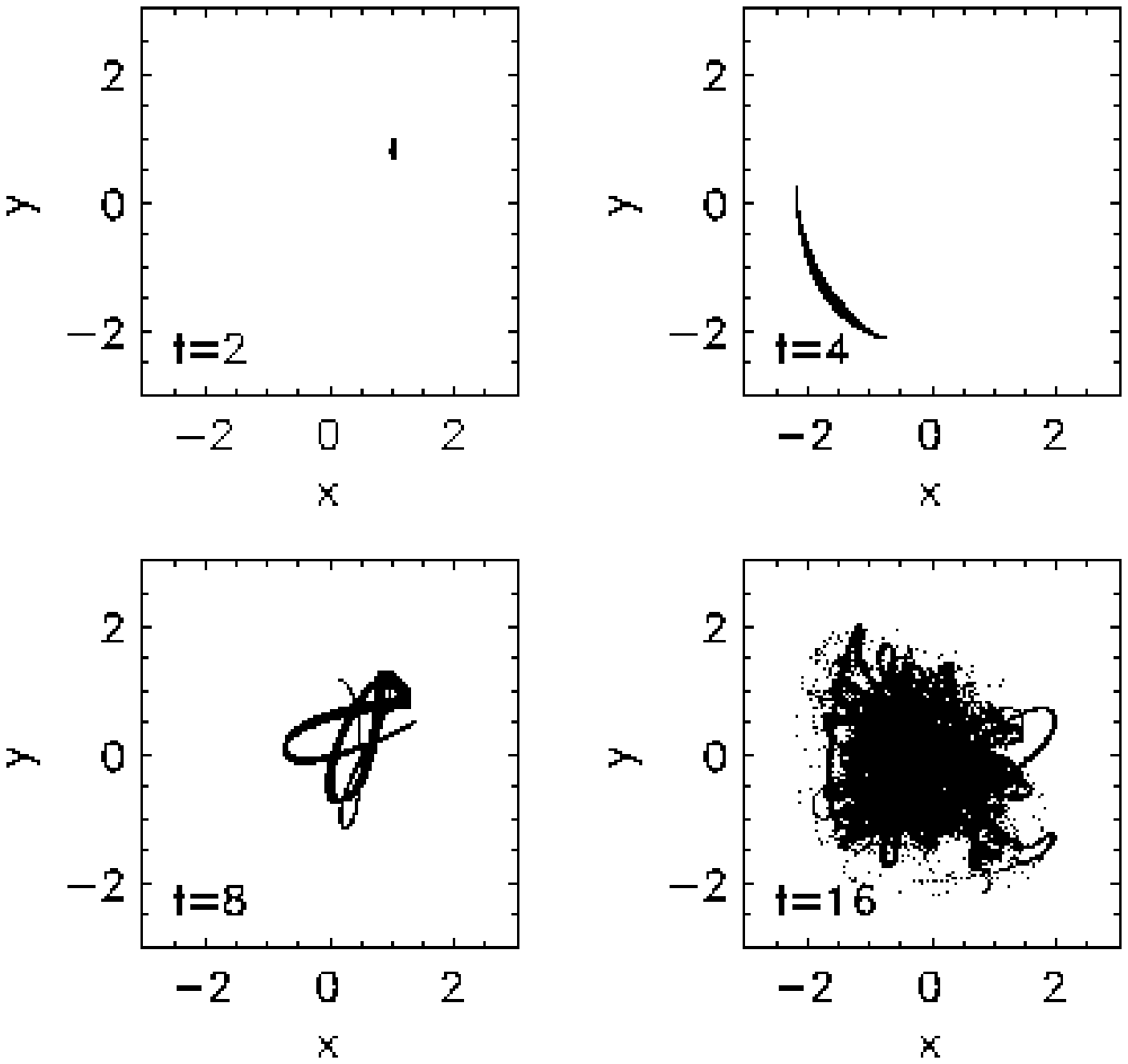}
           }
        \begin{minipage}{12cm}
        \end{minipage}
        \vskip -0.2in\hskip -0.0in
\caption{The evolution of an initially localised ensemble of 50,625 chaotic
orbits in the dihedral potential recorded at times $t=2,4,8,$ and $16$.}
\vspace{-0.2cm}
\end{figure}
\vfill\eject

\pagestyle{empty}
\begin{figure}[t]
\centering
\centerline{
        \epsfxsize=9cm
        \epsffile{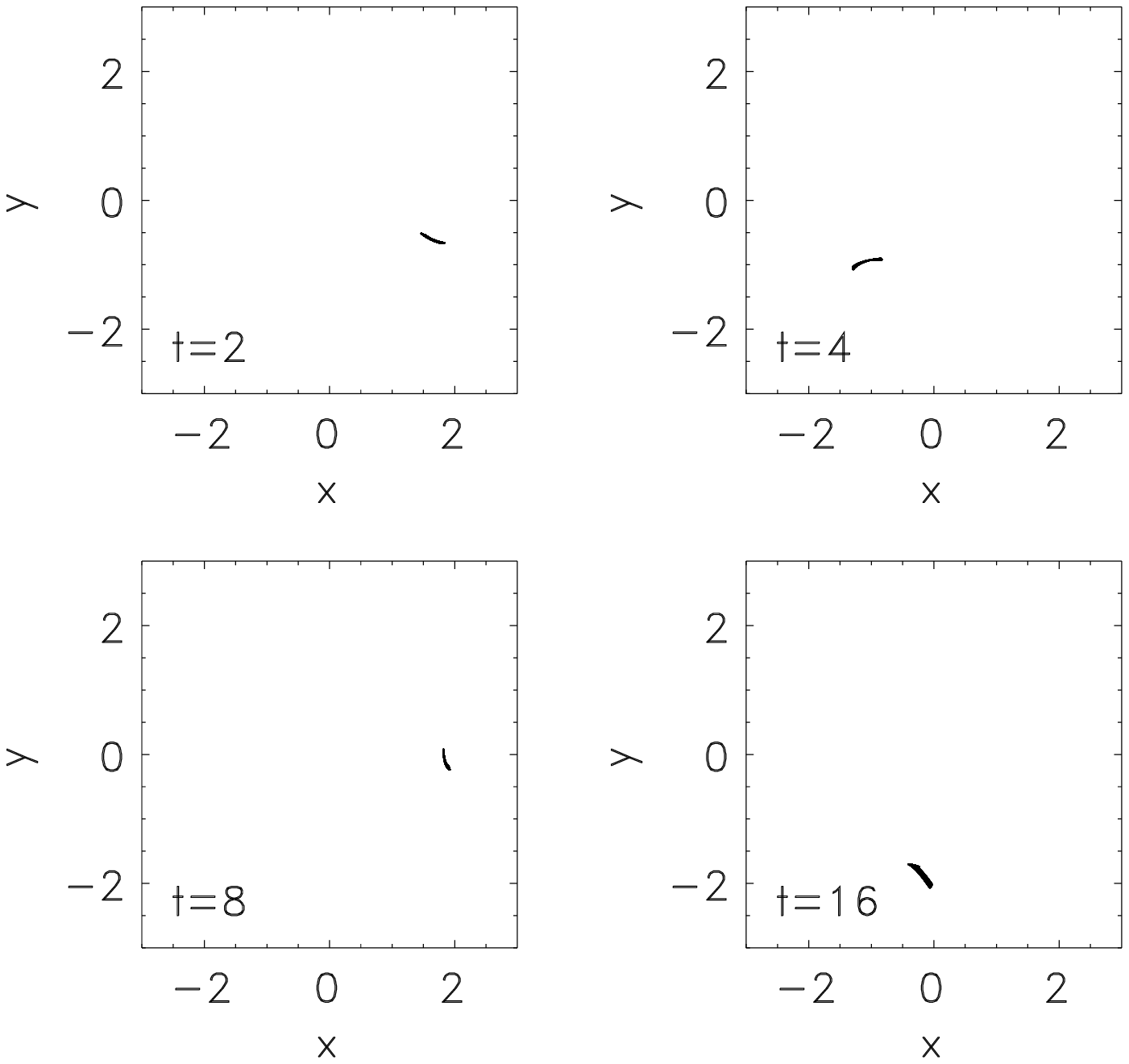}
           }
        \begin{minipage}{12cm}
        \end{minipage}
        \vskip -0.2in\hskip -0.0in
\caption{The evolution of an initially localised ensemble of 50,625 regular
orbits in the dihedral potential recorded at times $t=2,4,8,$ and $16$.}
\end{figure}
\vfill\eject

\end{document}